\documentclass{jetpl}
\usepackage{color,amsmath,amsfonts,graphics,epsfig,amssymb}
\usepackage[english]{babel}
\usepackage{hyperref}
\twocolumn
\lat
\usepackage{comment}

\title{
\begin{flushright}
\normalsize \rm INR-TH-2023-004
\end{flushright}
\vspace{2mm}
 Light-shining-through-wall cavity setups for probing ALPs 
}

\rtitle{ Light-shining-through-wall cavity setups for probing ALPs}

\sodtitle{ Light-shining-through-wall cavity setups for probing ALPs}

\author{Dmitry~Salnikov$^{*+}$
\thanks{salnikov.dv16@physics.msu.ru},
Petr~Satunin$^{*+}$, Maxim~Fitkevich$^{*\dagger}$, D.~V.~Kirpichnikov$^{*}$
}
\address{$^*$Institute for Nuclear Research of the Russian Academy
of Sciences, 60th October Anniversary Prospect, 7a, 117312  Moscow, Russia}
\address{$^{+}$Lomonosov Moscow State University, Leninskiye Gory, 119991 Moscow, Russia}
\address{$^{\dagger}$Moscow Institute of Physics and Technology,  
Institutskiy per., 9, 141701 Dolgoprudny, Moscow Region, Russia}

\rauthor{Dmitry~Salnikov, Petr~Satunin, Maxim~Fitkevich, D.~V.~Kirpichnikov}

\sodauthor{Dmitry~Salnikov, Petr~Satunin, Maxim~Fitkevich, D.~V.~Kirpichnikov}

\abstract{  We discuss the aspects of axion-like-particles (ALPs) searches with Light-Shining-through-Wall (LSW) experimental 
  setups consisted of two radio-frequency cavities. We compare the efficiencies of four setups which involve the cavity pump modes and external magnetic fields. 
 Additionally, we discuss the sensitivity dependence  both on  the relative position of two cylindrical cavities  and on their radius-to-length ratio.}
\begin{document}
\maketitle
\noindent\textbf{Introduction. } 
Light feebly-interacting pseudoscalar particles appear in modern particle physics in  various 
ways. Originally, a pseudoscalar particle called an axion was proposed in late 1970s to explain the 
strong CP problem in quantum chromodynamics \cite{Peccei:1977hh, Peccei:1977ur}
\footnote{We note that recently other solutions  for strong CP problem have been proposed within QCD,
 without introducing new particles beyond the standard model~\cite{Ai:2020ptm,Nakamura:2021meh,Yamanaka:2022bfj,Yamanaka:2022vdt}.
}. More general 
axion-like-particles (ALPs) are motivated by the string theory and appear in its low-energy  phenomenological description~\cite{Svrcek:2006yi, Arvanitaki:2009fg, Visinelli:2018utg}. In addition to the motivation 
for the particle physics models, axions and ALPs are of a great interest in cosmology  because they 
could make up a significant fraction of the dark matter in the Universe~\cite{Preskill:1982cy, Abbott:1982af, Dine:1982ah}. 

The Lagrangian for interacting  ALPs and photons can be written as follows
\begin{equation}\label{Lagrangian}
\mathcal{L}=-\frac{1}{4} F_{\mu\nu}F^{\mu\nu}+\frac12\,\partial_\mu a\, \partial^\mu a - \frac12\,m_a^2 a^2 -\frac{1}{4}\,g_{a\gamma\gamma}\,a\,F_{\mu\nu}\tilde{F}^{\mu\nu}\;,
\end{equation}
where $F_{\mu\nu}$ is the electromagnetic tensor and 
$\tilde{F}^{\mu\nu}=\frac{1}{2}\epsilon^{\mu\nu\alpha\beta}F_{\alpha\beta}$ is its dual, $a$ is the 
ALP field of mass $m_a$ with dimensionful photon-axion coupling $g_{a\gamma\gamma}$. The natural system of units $\hbar = c = k_B = 1$ is used.
Generally, $m_a$ and $g_{a\gamma\gamma}$ are treated as independent parameters.


A popular strategy of ALP searches is related to the cosmological (dark 
matter) and astrophysical probing. These ALPs can be detected by ground-based haloscopes 
(detection of dark matter ALPs) and helioscopes (ALPs can be produced hypothetically in the Sun)~ 
\cite{CAST:2017uph} (see e.g.~\cite{Irastorza:2018dyq} for a recent review). 

Another approach to probing ALPs implies both their production and detection in a laboratory, and
usually called Light-Shining-through-Wall (LSW) experiments 
\cite{Sikivie:1983ip,Anselm:1985obz,VanBibber:1987rq, Hoogeveen:1992nq}. The LSW setups consist of 
two cavities separated by a non-transparent 
wall. ALPs are produced in the first cavity by interaction of electromagnetic field components. Generated ALPs can pass through the wall and convert back to photons in the 
detection cavity. High intensity of initial electromagnetic field and the resonant amplification for the signal inside the cavities are required  because of the extremely small coupling $g_{a\gamma\gamma}$. Two wavelength ranges of EM 
fields are applicable to LSW: the optical range  setup including high intensity lasers and the radio range setup consisting of radio frequency cavities with high quality factors. 
Both ideas were realised in the experiments, ALPS (optical)~\cite{Ehret:2010mh} and CROWS (radio)~\cite{Betz:2013dza}. These experiments set the bound 
$g_{a\gamma\gamma} \simeq 10^{-7}\,\mbox{GeV}^{-1}$ 
for a wide range of ALP masses. However, this bound is three orders of magnitude weaker than the CAST helioscope limit $g_{a\gamma \gamma}\lesssim 6\times 10^{-11}\, \mbox{GeV}^{-1}$, 
\cite{CAST:2017uph}. For the moment, the ALPS-II laser experiment \cite{Ortiz:2020tgs} is under construction and its projected sensitivity exceeds CAST level.

In addition, the LSW radio experiments aimed at the ALP searches are of a great interest~\cite{Berlin:2022hfx}. Recently, several proposals with LSW radio cavities  appeared in the literature including superconducting radio frequency (SRF)  cavities~\cite{Gao:2020anb,Salnikov:2020urr, Bogorad:2019pbu}. In this letter we 
compare different LSW cavity setups including modification of the CROWS \cite{Hoogeveen:1992nq,Betz:2013dza}. 
Specifically, we study four setups: 
(i) an electromagnetic pump mode plus static magnetic field in the emitter cavity, static magnetic field in the receiver cavity \cite{Hoogeveen:1992nq}, we specify that as  {\bf MF (RF) emitter  + M$^*$F (RF) receiver} setup; 
(ii) two electromagnetic pump modes in the emitter cavity; an electromagnetic pump mode in the receiver cavity \cite{Gao:2020anb};
(iii) two electromagnetic pump modes in the emitter cavity, static magnetic field in receiver cavity \cite{Salnikov:2020urr};
(iv) an electromagnetic pump mode plus static magnetic field in the emitter cavity,  an electromagnetic pump mode in the receiver cavity.

\begin{figure}[h!]\centering
\includegraphics[width=0.49\linewidth]{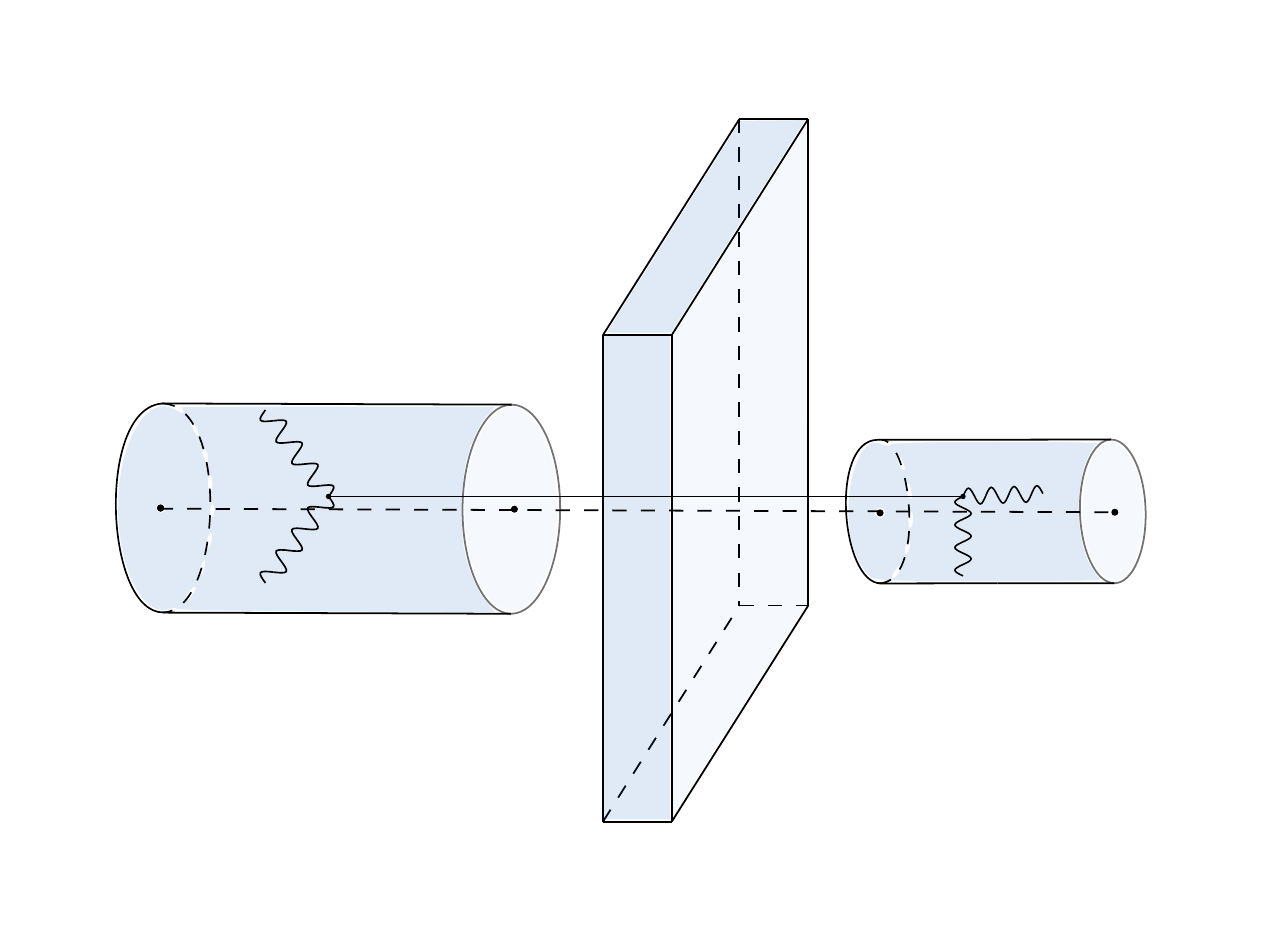}
\includegraphics[width=0.30\linewidth]{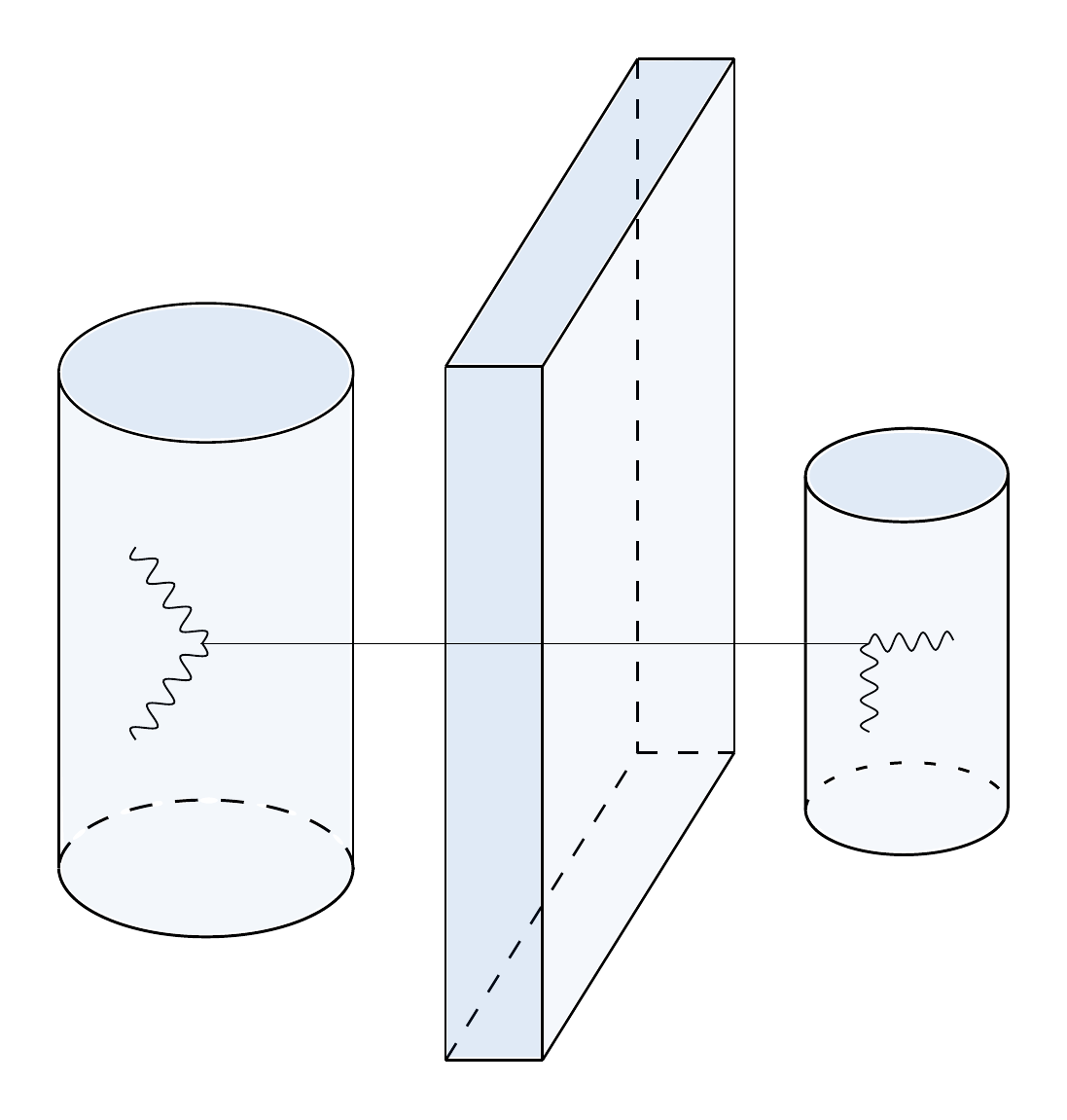}
\caption{Fig.~1: Two specific types of the experimental configuration consisting of two cylindrical cavities with 
(left panel) coaxial or (right panel) parallel orientation and screened by axion-penetrable wall. Wavy and 
solid lines represent electromagnetic field (cavity mode or magnetic field) and ALPs respectively.}
\label{fig:A}
\end{figure}

Another aspect of our analysis is geometry of the setup which can be adjusted in order to achieve higher sensitivity to ALPs parameters.
We study transfer of ALPs from the emitter to the receiver for all aforementioned  designs (i-iv) and discuss their optimal configuration, either coaxial or parallel 
(see e.~g.~Fig.~\ref{fig:A} for detail). 
Further, we investigate  $g_{a\gamma\gamma}$ sensitivity dependence  on the radius-to-length ratio of  production cylindrical cavity.  


 \vspace{0.3 cm}

\noindent\textbf{Axion electrodynamics.}
We briefly review the axion electrodynamics with the Lagrangian (\ref{Lagrangian}).
 The Euler-Lagrange equation 
for the ALP field reads,
\begin{equation}\label{Klein_Gordon_eqn1}
(\partial_\mu \partial^\mu +m_a^2)\,a=-\frac{1}{4}\, g_{a\gamma\gamma}\, F_{\mu\nu}\tilde{F}^{\mu\nu}\;,
\end{equation}
while the Maxwell's equations with an ALP-induced current read,
\begin{equation}
\label{Max_eqns1}
\partial_\mu F^{\mu\nu} = -g_{a\gamma\gamma}\,\partial_\mu a\,\tilde{F}^{\mu\nu}\;.
\end{equation}
One can rewrite Eqs.~\eqref{Klein_Gordon_eqn1} and \eqref{Max_eqns1} 
in terms of the electric and magnetic fields,
\begin{equation}\label{Klein_Gordon_eqn2}
(\partial_\mu \partial^\mu +m_a^2)\,a=g_{a\gamma\gamma}(\vec{E} \cdot \vec{B})\;,
\end{equation}
\begin{equation}\label{Maxwell1}
    (\vec{\nabla}\cdot \vec{E}) =  \rho_a\;, \quad\quad 
    [\vec{\nabla}\times \vec{B}] = \dot{\vec{E}} + \vec{j}_a\;,
\end{equation}
where the density of charge $\rho_a$ and current $\vec{j}_a$ are respectively given by 
\begin{equation}
\label{GeneralRhoaAndJa1}
    \rho_a = -g_{a\gamma\gamma} (\vec{\nabla}a\cdot \vec{B})\;, \quad\vec{j}_a = g_{a\gamma\gamma}( [\vec{\nabla}a\times \vec{E}] + \dot{a}\vec{B} )\;.
\end{equation}
Aforementioned equations describe  a way to produce ALPs by the electromagnetic field and the approach to detection of the ALP field in presence of background electromagnetic field. We further elaborate on this idea to compare the sensitivities of four types of the LSW setup for probing ALPs.

 \vspace{0.1 cm}

\noindent\textbf{The emitter cavity.} 
In this section, we consider the ALPs production. It is worth noting that Eq.~(\ref{Klein_Gordon_eqn2}) implies the invariant 
$F_{\mu\nu}\tilde{F}^{\mu\nu} = -4(\vec{E}\cdot\vec{B})$ should be non-zero in order to produce ALPs by the electromagnetic field. 
Following this requirement, we consider two options for the production of ALPs using RF cavities:

(i) a normally conducting RF cavity with a single pump mode with frequency $\omega_0$ immersed in a strong static magnetic field $\Vec{B}_{\rm ext}$. We use the notation {\bf MF emitter} (i.~e.~pump mode ({\bf M}) $+$ 
magnetic field ({\bf F})) for this case throughout the paper;

(ii) a superconducting RF cavity with two pump modes at frequencies $\omega_{1,2}$. We use notation {\bf MM emitter} (pump mode ({\bf M}) $+$ pump mode ({\bf M})) for this setup.

 It is worth mentioning, that in the steady regime both EM-source (r.h.s of Eq.~\ref{Klein_Gordon_eqn2}) and the induced axion field have the same frequency $\omega_a$.  
For the  MF emitter case, the source function in the Eq.~\eqref{Klein_Gordon_eqn2} contains a single component oscillating at  the frequency $\omega_a = \omega_0$.  However, for the MM emitter case, there are two components at frequencies $\omega_a = \omega_{\pm} = \omega_2 \pm \omega_1$ ($\omega_2 > \omega_1$). As a result, each particular 
combination of the field for both MF emitter and  MM emitter can be written in the general form,
\begin{equation}\label{Source}
    f(t,\vec{x}) = g_{a\gamma\gamma}E^{\rm em}_0 B^{\rm em}_0 \, {\rm Re} \, \left[(\vec{\cal E}\cdot \vec{\cal B})(\vec{x}) \cdot e^{-i\omega_a t} \right]\;,
\end{equation}
where $E^{\rm em}_0, B^{\rm em}_0$  are typical values of the emitter EM fields, $(\vec{\cal E}\cdot \vec{\cal B})(\vec{x})$ is a dimensionless function determined by the production approach.

For the RF cavity that  emits the ALPs (MF emitter), $E_0^{\rm em}$ is a typical amplitude of pump mode taken on the cavity wall\footnote{The peak amplitude is greater by the factor of ${\cal O}(1)$ and depends on the particular mode and cavity geometry.}, $B_0^{\rm em}$ is a magnitude of the static magnetic field, and
the typical combination of the normalized  fields in Eq.~(\ref{Source}) 
can be  written as  follows
\begin{equation}
\label{dimFactMFemitter}
   (\vec{\cal E}\cdot \vec{\cal B})(\vec{x}) = (\vec{\cal E}_0(\vec{x}) \cdot \vec{\cal B}_{\rm ext})\;,
\end{equation}
where $\Vec{\cal E}_0(\vec{x})$ is a dimensionless electric field of the pump mode,
and $\vec{\cal B}_{\rm ext}$ is a unit vector that is collinear to  the  
magnetic field direction.

For the MM emitter case, $E_0^{\rm em}$ and $B_0^{\rm em}$ represent the surface electric and magnetic fields amplitudes of both pump modes
and the dimensionless functions in Eq.~\eqref{Source} can be written for $\omega_a = \omega_+$ and $\omega_a=\omega_-$ respectively as follows 
\begin{align}
  &  (\vec{\cal E}\! \cdot \! \vec{\cal B})_{+}(\vec{x}) = \! \dfrac{1}{2} \!\left[ (\vec{\cal E}_1(\vec{x})\! \cdot\! \vec{\cal B}_2(\vec{x}))\!  + \! (\vec{\cal E}_2(\vec{x})\! \cdot\! \vec{\cal B}_1(\vec{x}))\right],\\
 &    (\vec{\cal E}\! \cdot\!  \vec{\cal B})_{-}(\vec{x})=\! \dfrac{1}{2}\!\left[ (\vec{\cal E}^*_1(\vec{x})\! \cdot\! \vec{\cal B}_2(\vec{x}))\!  +\!  (\vec{\cal E}_2(\vec{x})\! \cdot\! \vec{\cal B}^*_1(\vec{x}))\right],
\end{align}

where $\vec{\cal E}_i(\vec{x})$, $\vec{\cal B}_i(\vec{x})$, $i=1,2$, are dimensionless electric and magnetic fields of pump modes.

Eq.~\eqref{Klein_Gordon_eqn2} implies the following solution,
\begin{align}
\label{Produced_ALPs}
   a(t,\vec{x}) = g_{a\gamma\gamma}\, E^{\rm em}_0 B^{\rm em}_0 \int_{V_\mathrm{em}} d^3 x'\;\mathrm{Re}\Bigl[(\vec{\cal E} \cdot \vec{\cal B})(\vec{x}')  
   \nonumber
   \\    \times 
   \dfrac{e^{ik_a|\vec{x} - \vec{x}'| - i\omega_a t}}{4\pi|\vec{x} - \vec{x}'|} \Bigr] \equiv  \mathrm{Re}\left[ a(\vec{x})\,e^{-i \omega_a t} \right]\;, 
\end{align}
where $k_a=\sqrt{\omega_a^2 - m_a^2}$ are typical momenta of the produced ALPs, integration is performed over the emitter volume $V_{\rm em}$. 
For MF emitter,  the dimensionless factor in Eq.~(\ref{Produced_ALPs})  is defined by 
Eq.~(\ref{dimFactMFemitter}). However, for  MM emitter the solution of  
Eq.~\eqref{Klein_Gordon_eqn2} splits into two frequency components $\omega_\pm$. 
 Note that the produced ALPs of frequency component $\omega_-$ is at least an order of magnitude smaller than $\omega_+$ component \cite{Salnikov:2020urr}, so further we deal with $\omega_+$ only.
One can replace formally $i k_a$ with $-\kappa_a=-\sqrt{m_a^2 - \omega_a^2}$ in Eq.~(\ref{Produced_ALPs}) for the relatively heavy ALP mass limit $m_a \gtrsim \omega_a$.  Then the ALP field amplitude decreases exponentially 
as  $a(\vec{x}) \propto \exp{\left(-\kappa_a |\vec{x}| \right)}/|\vec{x}|$ outside  the production 
cavity.  Therefore,   the detecting cavity signal is suppressed for $m_{a} \gtrsim \omega_a$ mass range.

 \vspace{0.3 cm}

\noindent\textbf{The receiver cavity.}
In this section, we investigate the issue of the ALP signal  detection in the receiver cavity. 
A resonant generation of electromagnetic modes in the detecting cavity caused by the axion-induced current,
$j^\nu_a = (\rho_a, \vec{j}_a)$ where the density of the
effective charge $\rho_a(\vec{x},t)$ and the effective  current $\vec{j}_a(\vec{x},t)$ are given by Eqs.~(\ref{GeneralRhoaAndJa1}). 
 Two options are assumed for detection: 

(i) the receiver cavity is a normally conducting one, and it is immersed into external constant 
    magnetic field $\vec{B}_{\rm ext}$. 
    We use the notation {\bf M$^*$F receiver} (induced signal 
    mode ({\bf M$^*$}) + magnetic field ({\bf F}) of the receiver) for that case (the label {\bf M}$^*$ 
    denotes the mode that we expect to detect throughout the paper);
   
    (ii) the receiver cavity is superconducting, and it  is pumped by the detecting mode.
    We use the notation {\bf M$^*$M receiver} 
    (induced signal mode ({\bf M$^*$}) + pump mode ({\bf M})  of the receiver) for this setup of the cavity.

One can show (see e.~g.~Ref.~\cite{Berlin:2021txa} and references therein for detail)
that generating field is a combination of solenoidal and potential modes, however only the solenoidal modes can be resonantly enhanced. The typical magnitude of the signal
 can be characterized by the expression~\cite{Gao:2020anb, Bogorad:2019pbu}
\begin{equation}\label{Amplitude}
    G = -\dfrac{Q_{\rm rec}}{\omega_s}\cdot \dfrac{1}{V_{\rm rec}}\int_{V_{\rm rec}} d^3 x\, (\vec{\cal E}^*_s\cdot  \vec{j}_a)\;,
\end{equation}
where $Q_{\rm rec}$ is a quality factor for the receiver eigenmode that depends on the electric field near the cavity walls and corresponding power losses due to non-linearities (see e.~g.~Ref.~\cite{Berlin:2019ahk} and references therein for details), $V_{\rm rec}$ is the volume of the receiver cavity,  $\omega_s$ is a frequency of the receiver signal eigenmode, and $\Vec{\cal E}_s(\vec{x})$ is a dimensionless signal
eigenmode that is normalized as follows \cite{Bogorad:2019pbu}
\begin{equation}\label{Volume_normalization}
    \int\limits_{V_{\rm rec}} d^3 x \, |\vec{\cal E}_s(\vec{x})|^2 = V_{\rm rec}\;. 
\end{equation}

The specific form of the current $\vec{j}_a$ in the Eq.~\eqref{Amplitude} depends on the way of 
ALP detection. It is  remarkable that the general  expression of the overlapping integral in Eq.~(\ref{Amplitude}) for both M$^*$F and M$^*$M receivers 
can be written in the following explicit form
\begin{equation}\label{Detecting_Integral}
    \int\limits_{V_{\rm rec}}\! \!\! d^3 x\, (\vec{\cal E}^*_s \!  \cdot \!  \vec{j}_a) = -i\omega_s g_{a\gamma\gamma} B^{\rm rec}_0\! \!\! \int\limits_{V_{\rm rec}}\! \! \!d^3 x\,  (\vec{\cal E} \!  \cdot \!  \vec{\cal B})^*\!(\vec{x}) a(\vec{x})\;,
\end{equation}
where $B^{\rm rec}_0$ is a characteristic magnetic field of the detection cavity and 
$(\vec{\cal E}\cdot \vec{\cal B})^*(\vec{x})$ is a dimensionless complex-conjugated function that is associated with a specific way of ALP detection. 

More specifically, for the M$^*$F receiver, $B_0^{\rm rec}$ is the value of the external magnetic field, and dimensionless function has the following form
\begin{equation}
   (\vec{\cal E}\cdot \vec{\cal B})^*(\vec{x}) = (\vec{\cal E}_s(\vec{x}) \cdot \vec{\cal B}_{\rm ext})^* \; ,
\end{equation}
where $\vec{\cal B}_{\rm ext}$ is a unit vector co-directed with a magnetic field inside the receiver cavity.

For the M$^*$M receiver, $B_0^{\rm rec}$ is the magnetic field amplitude of the pump mode and the combinations of dimensionless functions are given by 
\begin{align}
    (\vec{\cal E} \!\cdot\! \vec{\cal B})^*_{+}(\vec{x}) & \!=\! \dfrac{1}{2}\!\left[ (\vec{\cal E}_s(\vec{x})\!\cdot\!\vec{\cal B}_d(\vec{x})) + (\vec{\cal E}_d(\vec{x})\!\cdot\!\vec{\cal B}_s(\vec{x}))\right]^*,\\
    (\vec{\cal E}\!\cdot\! \vec{\cal B})^*_{-}(\vec{x}) & \!= \!\dfrac{1}{2}\!\left[ (\vec{\cal E}_s(\vec{x})\!\cdot\!\vec{\cal B}^*_d(\vec{x})) \!+\! (\vec{\cal E}^*_d(\vec{x})\!\cdot\!\vec{\cal B}_s(\vec{x}))\right]^* \; ,
\end{align}
for $\omega_a =  \omega_s + \omega_d$ and $ \omega_a = \omega_s- \omega_d \ (\omega_s > \omega_d)$ respectively, 
here $\omega_d$ is a receiver pump mode frequency,  $\vec{\cal E}_{s(d)}(\vec{x})$ and $\vec{\cal B}_{s(d)}(\vec{x})$  are  dimensionless electric and magnetic fields respectively for the signal mode (detection pump mode in Ref.~\cite{Gao:2020anb}). 

 \vspace{0.3 cm}

\noindent\textbf{Signal power.}
Here we discuss the signal induced by the axion field for the cavity
experimental setups. 
To be more concrete,  by using Eqs.~(\ref{Produced_ALPs}) 
and (\ref{Detecting_Integral}) we can  rewrite  the amplitude  in Eq.~\eqref{Amplitude} in general form
\begin{equation}
    G = iQ_{\rm rec}g^2_{a\gamma\gamma}E^{\rm em}_{0}B^{\rm em}_{0}B^{\rm rec}_{0}\cdot \dfrac{V_{\rm em}V_{\rm rec} {\cal G}}{\Delta}\;,
    \label{GfactorDefinition2}
\end{equation}
where $\Delta$ is typical distance between cavities, and the dimensionless factor ${\cal G}$
is given by the following expression
\begin{equation}
    {\cal G}\, = \!\!\! \int\limits_{V_{\rm rec}}\! \!\! \dfrac{d^3 x}{V_{\rm rec}} \!\!\int\limits_{V_{\rm em}} \!\! \dfrac{d^3x'}{V_{\rm em}} \, (\vec{\cal E}\!\cdot\! \vec{\cal B})^*(\vec{x}) \,  (\vec{\cal E}\!\cdot\! \vec{\cal B})(\vec{x}')\, \dfrac{e^{ik_a|\Vec{x} - \vec{x}'|}}{4\pi} \dfrac{\Delta}{|\vec{x} - \vec{x}'|}.
\end{equation}
In the steady regime the averaged signal power can be expressed in the following form, 
\begin{equation}\label{Signal_Power}
    P_{\rm signal} \! = \! \dfrac{\omega_s}{Q_{\rm rec}} \!\! \int\limits_{V_{\rm rec}} \!\! d^3 x \, \langle |\vec{E}_s^2(\vec{x},t)| \rangle_t \!= \!\dfrac{\omega_s}{Q_{\rm rec}} \dfrac{1}{2} |G|^2 V_{\rm rec},
\end{equation}where $\vec{E}_s(\vec{x},t)$  is a signal solenoidal electric field that is resonantly enhanced by the ALP in the receiver.  It 
is important to note  that Eq.~(\ref{Signal_Power}) implies 
$\langle |\vec{E}(\vec{x},t)|^2 \rangle_t = \langle |\vec{B}(\vec{x},t)|^2 \rangle_t$ 
(see e.~g.~Ref.~\cite{Berlin:2021txa} and references therein for detail). 
We imply that time averaging of the squared magnitude results  in the replacement $\langle ... \rangle_t \to  1/2$~\cite{Bogorad:2019pbu}. We note that this approach  provides the same result as the power spectral density calculation~\cite{Gao:2020anb,Berlin:2019ahk} for the narrow signal bandwidth limit.

We estimate sensitivity numerically as maximum output in the receiver cavity that is  given by the Dicke radiometer equation,
\begin{equation}
 \mathrm{SNR} =\frac{P_{\mathrm{signal}}}{P_\mathrm{noise}}\cdot  \sqrt{t\Delta \nu}\;,
 \label{SNR1}
\end{equation}
where $t$ is an integration time for a signal, $\Delta \nu$ is its bandwidth and $P_{\rm noise}$ is a power of thermal 
noise which can be estimated as $P_{\rm noise}\simeq T\Delta \nu$ in the limit $\omega_s \ll T$, where $T\simeq 1.5\,\mbox{K}$ is the typical 
temperature of the receiver. We consider two 
options for $\Delta \nu$: the bandwidth of a cavity mode itself (i.e. $\Delta \nu \simeq \nu_s/Q_{\rm rec}$, where 
$\nu_s=\omega_s/(2\pi)$) and the narrowest possible bandwidth of a pump generator, which can be as small as 
$\Delta \nu  \simeq 1/t$ (see e.~g.~Refs.~\cite{Bogorad:2019pbu, Gao:2020anb} and references therein). 

It is worth noticing that in the present paper we study only noise from the thermal 
fluctuation. This implies that the other sources of the background should be significantly 
mitigated. The latter includes the mechanical noise and oscillator phase noise  that has been considered 
explicitly in Refs.~\cite{Berlin:2019ahk,Gao:2020anb}.
However we conservatively expect that these backgrounds can be subdominant to the the thermal noise by 
further  optimisation of the experimental facility.  

Finally, by using Eqs.~(\ref{Signal_Power}) and (\ref{SNR1}) one can obtain the general  formula for the expected sensitivity,
\begin{equation}
\label{master}
    g_{a\gamma\gamma}\!\! =\!\! \left[\! \dfrac{2 \Delta^2 T \, {\rm SNR}  }{\omega_sQ_{\rm rec}E^2_{0,{\rm em}}\!B^2_{0,{\rm em}}\!B^2_{0,{\rm rec}}\!V^2_{\rm em}\!V_{\rm rec}{|\cal G}|^2}\! \right]^{\!\!\frac{1}{4}} \!\!\!\! \left(\!\dfrac{\Delta \nu}{t}\!\right)^{\!\frac{1}{8}}\!\!\!\!,
\end{equation}
which is valid for estimation of the ALP expected limit (${\rm SNR} \simeq 5$)
for all benchmark setups considered in the next section. 

\begin{figure}[h!]\centering
\includegraphics[width=0.9\linewidth]{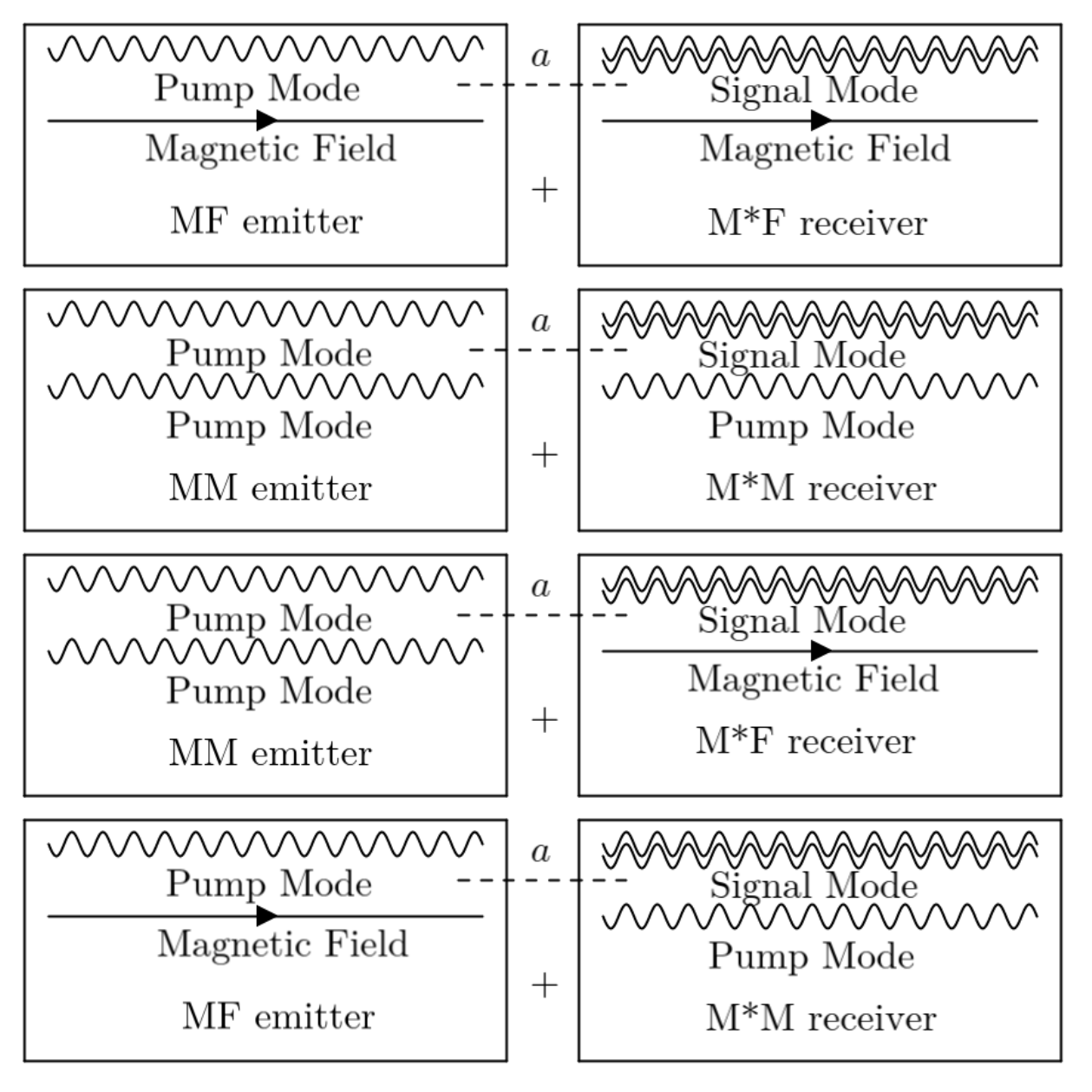}
\caption{Fig.~2: Four general types of the experimental proposals.}
\label{fig:TypicalSetup}
\end{figure}

 \vspace{0.3 cm}

\noindent\textbf{The expected reach.}
Now we compare the efficiencies of   four different experimental setups for probing 
ALPs with LSW methods: (\text{i})  {\bf MF (RF) emitter  + M$^*$F (RF) receiver};  (\text{ii}) 
{\bf MM (SRF) emitter + M$^*$M (SRF) receiver};  (\text{iii}) {\bf  MM (SRF) emitter + M$^*$F
(RF) receiver}; (\text{iv}) {\bf  MF (RF) emitter + M$^*$M
(SRF) receiver} (for details see e.~g.~Fig.~\ref{fig:TypicalSetup}). 
In addition, we study in detail the sensitivity dependence on the
spatial geometry (relative position of emitter and  receiver cylindrical cavities) and 
radius to length  ratio  $R/L$  of  the ALP emitter for these benchmark  experimental  proposals. 

\vspace{0.3 cm}

\begin{figure*}[h!]\centering
\includegraphics[width=0.4\linewidth]{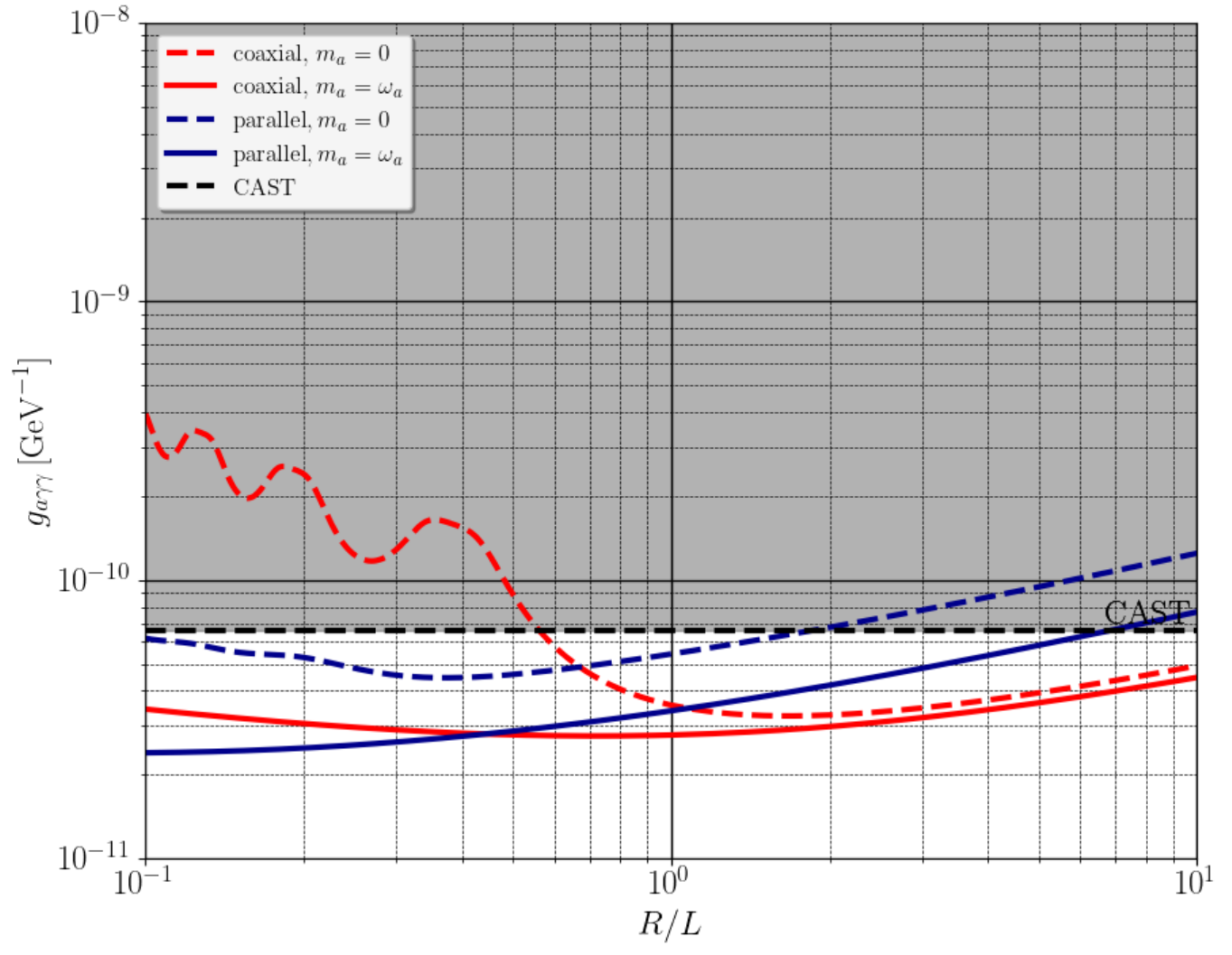}
\includegraphics[width=0.4\linewidth]{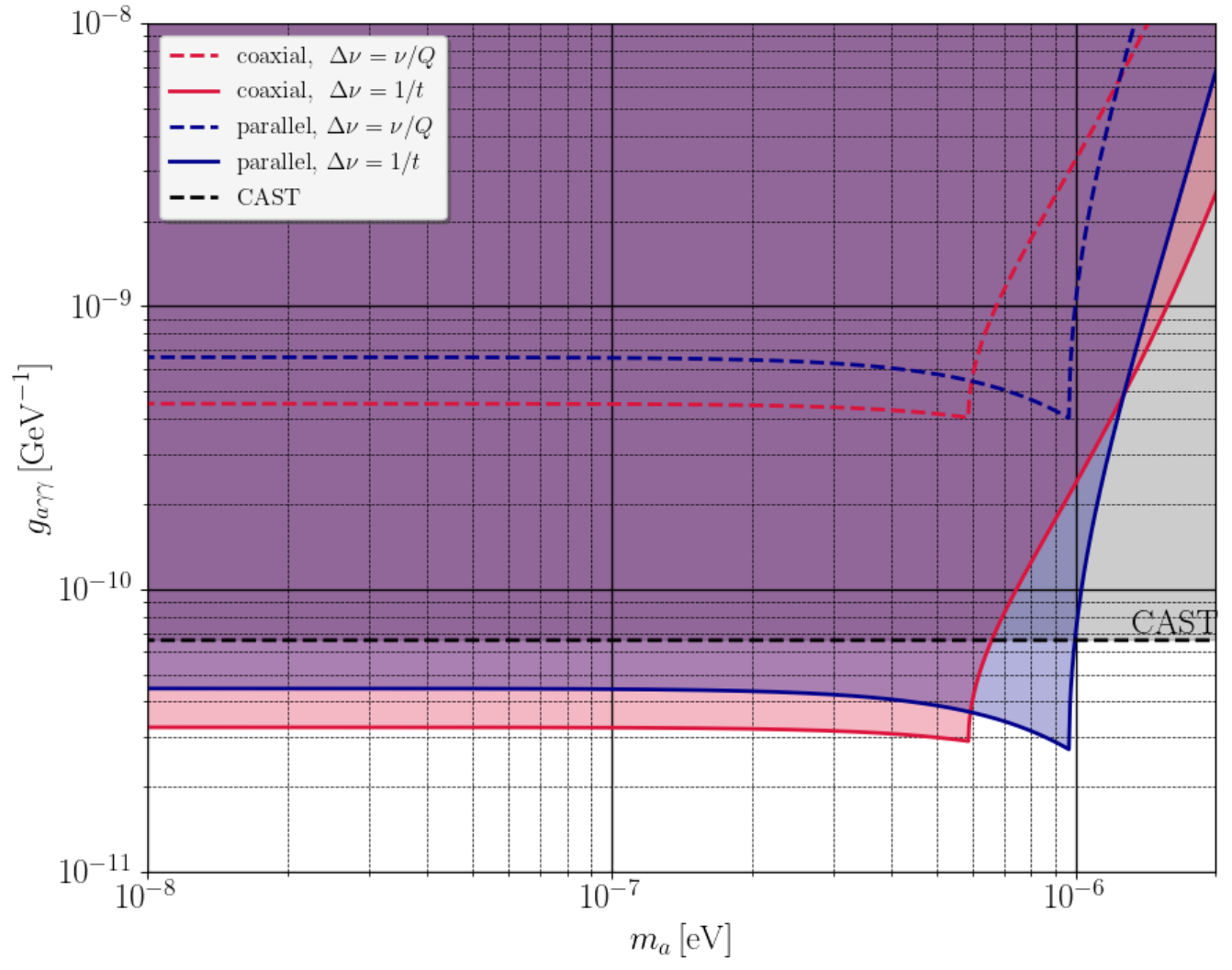}
\caption{Fig.~3: The  sensitivity of  {\bf MF emitter + M$^*$F receiver}  setup for both coaxial and  parallel cavity locations and the TM$_{010}$ emitter and receiver 
modes. Left  panel: the dependence on the emitter cavity radius-to-length ratio  $R/L$ for the typical volume $V_{\rm em} = 1 \, {\rm m}^3$. 
Right panel: expected reach as a function of ALPs mass at optimal $R/L$ 
for coaxial ($R/L \simeq 1.67$, $R \simeq 0.81 \, {\rm m}$, $L \simeq 0.49 \, {\rm m}$) and parallel ($R/L \simeq 0.37$, $R \simeq 0.49 \, {\rm m}$, $L \simeq 1.32 \, {\rm m}$) geometries. The distance between cavity walls is $\Delta = 0.5$ m,  the cavity volumes are $V_{\rm rec}= V_{\rm em}=1$ m$^3$. The integration time is $t = 10^6 \ {\rm s}$. The temperature of the receiver is taken as $T=1.5$ K.}
\label{fig:CROWS}
\end{figure*}

\noindent\textbf{MF  emitter + M$^*$F  receiver.}
 At first let us consider the typical LSW setup  consisting of two RF cavities  which are placed both into a strong static magnetic field~\cite{Hoogeveen:1992nq}. 
 The experimental realization of that idea was carried out by the CROWS experiment \cite{Betz:2013dza}. 
 We show the sensitivity of this type of experiment for the characteristic volume of the emitter 
 and receiver cavities $V_{\rm rec} = V_{\rm em} \simeq 1\, \mbox{m}^3$, the receiver quality 
 factor\footnote{Note that the  receiver quality factor can be as large as  $Q_\text{rec} \sim 10^6$ 
 if one exploits specific superconducting tapes in relatively small cavity 
 volumes \cite{PhysRevApplied.17.L061005}. We expect that it would be a challenging issue for larger cavity   volumes.}
 $Q_{\rm rec} \simeq 10^5$. 
 We consider the characteristic magnitude of the emitter pump mode
 $E^{\rm em}_{0} = 3 \, {\rm MV/m} \ (B^{\rm em}_0 = 0.01 \, \rm{T})$. It is worth noticing that 
 the emitter quality factor $Q_{\rm em}$ can be smaller by several times  \cite{Tantawi:2007zzb} 
 than the receiver quality factor $Q_{\rm rec} \simeq 10^5$  due to the high power of the pump 
 mode. Given the volume and quality factor, the emitter power is of the order of $P_{\rm em} \sim 100 \, \rm{kW}$. The latter to be a reasonable power for  injection in the emitter.
  The typical value of the static magnetic field are taken as $B_0^{\rm em} = B_0^{\rm rec} = 3 \, {\rm T}$ in Fig.~\ref{fig:CROWS}. 
  The distance between receiver and emitter  walls is $\Delta = 0.5$ m. The pump mode of the emitter and the signal mode of the receiver are TM$_{010}$.

 In Fig.~\ref{fig:CROWS} (left panel)  we show the expected sensitivity of the 
 setup as a function of $R/L$ for both parallel and coaxial designs of the cavities (see 
 e.~g.~Fig.~\ref{fig:A} for detail),  
 we also set the ALP benchmark masses to be $m_{a}=0$ and $m_{a} \simeq \omega_a$. We take into account that the 
 $Q_{\rm rec}$  depends on the chosen mode and cavity geometry, 
 $Q_{\rm rec} =  {\omega_s}/R_s\,\cdot\, V_{\rm rec}/S_{\rm rec} \propto  \sqrt{\omega_s}\, V_{\rm rec}/S_{\rm rec}$, where $R_s$ is the surface resistance and $S_{\rm rec}$ is the surface area of the 
 detector cavity \cite{Hill2014}, here we also set the value $Q_{\rm rec} = 10^5$ for $R/L=1$. 
 We consider sensitivity for $m_a = 0$ as the most important setup characteristic 
 compared to the resonant bound at $m_a = \omega_s$ regime throughout the paper. It implies that the typical bounds at $m_a = 0$  cover the larger logarithmic mass scale range ($m_a \lesssim \omega_a/2$) in $(g_{a\gamma\gamma}, m_a)$ plane.    
 It turns out that 
 coaxial design  for $R/L \gtrsim 1$ is more preferable. It is remarkable that in this case the typical expected reaches  for  both masses $m_a = 0$ and
 $m_a\simeq \omega_a$ coincide by the order of the magnitude at the level of 
 $g_{a\gamma \gamma} \simeq 3\times 10^{-11}\; \mbox{GeV}^{-1}$. However, there is a notable difference between 
 the  expected reaches at $m_a = 0$ and $m_a \simeq \omega_s$ for  the parallel design. Note that optimal 
 radius to length ratio (that implies better sensitivity on $g_{a\gamma\gamma}$ in case of $m_a = 0$)
is   $R/L \simeq 1.67$ for coaxial design and $R/L \simeq 0.37$ for parallel design.
 
 In Fig.~\ref{fig:CROWS} (right panel) we show the expected reach as a function of the ALP mass $m_a$ 
 for both coaxial and parallel  locations of  the cavities  at the optimal ratios $R/L$ assuming two options of the signal bandwidth $\Delta \nu \simeq \nu/Q_{\rm rec}$ and $\Delta \nu \simeq 1/t$, where $t\simeq 10^6 \, 
 \mbox{s}$ is the typical time of measurement. The conservative cavity bandwidth
 $\Delta \nu \simeq \nu/Q_{\rm rec}$  yields the expected limit $g_{a\gamma\gamma}  \lesssim 5\times 10^{-10}\, \mbox{GeV}^{-1}$ that is 
 weaker than the CAST constraint~\cite{CAST:2017uph}. However, the optimistic bandwidth $\Delta \nu \simeq 1/t$ can  provide the expected reach $g_{a\gamma\gamma}\lesssim 3 \times 10^{-11}\, \mbox{GeV}^{-1}$ for $m_{a} \lesssim \omega_a/2$.
 

It is worth noting that corresponding heat production makes it
challenging to keep the emitter temperature at a required level.
On the other hand, relatively  small temperatures $T \lesssim 1\, \mbox{K}$ are required for the receiver cavity in order to avoid the problem of thermal noise. 
So that keeping two cavities at different temperatures  would require a development of sophisticated RF methods for the regarding setup. 


 \vspace{0.3 cm}

\noindent\textbf{MM emitter + M$^*$M receiver.}
\begin{figure*}[h!]\centering
\includegraphics[width=0.4\linewidth]{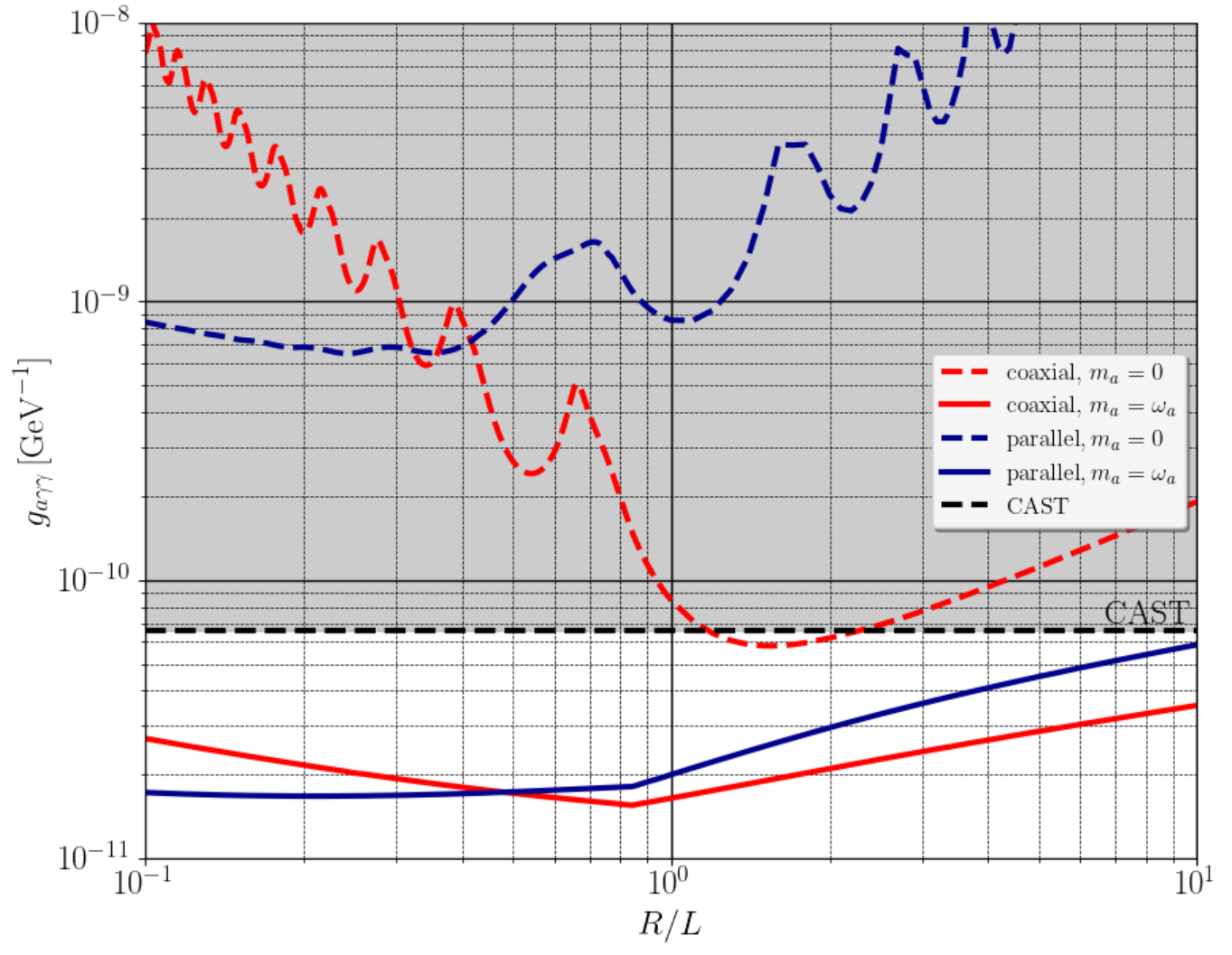}
\includegraphics[width=0.4\linewidth]{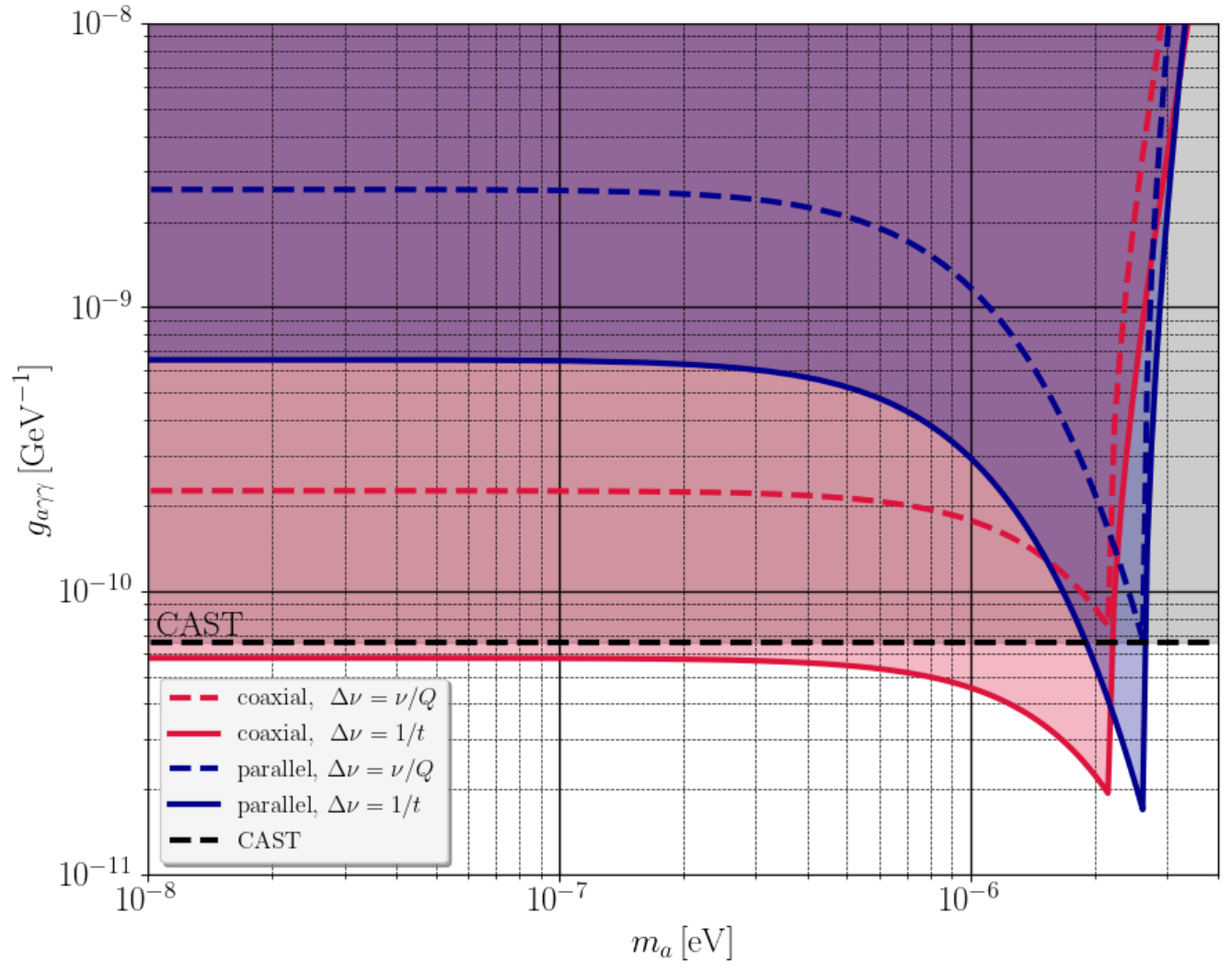}
\caption{Fig.~4: The sensitivity of {\bf MM emitter + M$^*$M receiver}
cavity setup. This facility implies combination of TM$_{010}$ + TE$_{011}$ production pump modes. 
The pump mode of a receiver and its signal mode are chosen to be TM$_{010}$ and TE$_{011}$ respectively. The case of ALPs frequency $\omega_a = \omega_+$ is considered.
Left panel:  the expected limit $g_{a\gamma\gamma}$ as a function of  production 
cavity radius-to-length ratio $R/L$ (we set the emitter volume at $V_{\rm em} = 1 \, {\rm m}^3$). 
Right panel: Sensitivity as a 
function of ALPs mass at optimal $R/L$  for coaxial 
($R/L \simeq 1.60$, $R \simeq 0.80 \, {\rm m}$, $L \simeq 0.50 \, {\rm m}$) and parallel 
($R/L \simeq 0.35$,  $R \simeq 0.48 \, {\rm m}$, $L \simeq 1.37 \, {\rm m}$) designs.  
The distance between cavity walls  is $\Delta = 0.5 \, \mbox{m}$, 
the volume of each cavity is $V_{\rm em} = V_{\rm rec}=1$ m$^3$. 
Integration time is $t = 10^6 \, {\rm s}.$ The temperature of the receiver is taken as $T=1.5$ K.}.
\label{fig:GAO}
\end{figure*}
The second setup of our interest  consists of two equal SRF cavities~\cite{Gao:2020anb}. In the emitter 
cavity, ALPs are generated by an interaction of two cavity modes. In the detection cavity, produced ALPs 
interact with a single pump mode (which coincides with one of the production cavity pump modes), producing the 
resonantly enhanced signal mode in the receiver cavity. The magnitude of the surface amplitude of pump modes for an SRF cavity
to be as small as $B^{\rm em, rec}_0 \lesssim 0.1 \, {\rm T} \ (E_0^{\rm em, rec} \lesssim 30 \, {\rm MV/m})$ to avoid the 
superconductivity state destruction. The volume of the emitter and receiver cavities 
$V_{\rm rec} = V_{\rm em} \simeq 1\, \mbox{m}^3$, their  quality factor 
 $Q \simeq 10^{10}$.  This high quality factor implies specific fine tuning of the emitter cavity frequency, see \cite{Romanenko:2023irv}. The expected power of the emitter cavity is $P_{\rm em} \simeq 0.1 \, {\rm kW}$. In Fig.~\ref{fig:GAO}
the typical sensitivities for the regarding LSW setup are presented. 

In Fig.~\ref{fig:GAO} (left panel)
the expected reach as function of emitter radius-to-length ratio $R/L$ is shown. As in previous case, we take into account the dependence of the quality factor $Q$ on cavity geometry which reads $Q_{\rm rec} = \omega_s/R_s \cdot V_{\rm rec}/S_{\rm rec} \propto {\omega_s}^{-1}\, V_{\rm rec}/S_{\rm rec}$ for SRF cavity \cite{1062561}, and fix the value as $Q_{\rm rec} = 10^{10}$ for $R/L=1$.
It turns out that the 
optimal magnitude of $R/L$ for the coaxial cavity location and for the ALP  mass limit 
$m_a=0$ is  $R/L\simeq 1.6$. The regarding expected sensitivity is $g_{a\gamma\gamma} \lesssim 5\times 10^{-11}\, 
\mbox{GeV}^{-1}$ that is comparable with the CAST bound $g_{a\gamma \gamma}\lesssim 6\times 10^{-11}\, \mbox{GeV}^{-1}$.
For parallel location of the cavities,
the optimal radius-to-length ratio is  $R/L\simeq 0.35$ implying $m_a =0$. 
 We note that zero axion mass bounds  
$g_{a\gamma\gamma } \lesssim  6\times 10^{-10}\, \mbox{GeV}^{-1}$ are ruled out by the CAST.  
The signal cavity bandwidth is  chosen to be at the level 
$\Delta \nu \simeq 1/t$, where $t\simeq 10^6 \, \mbox{s}$ is a typical  time of the measurements. 

In Fig.~\ref{fig:GAO} (right panel) we show the expected limit $g_{a\gamma\gamma}$ of  this setup as a function of the ALP mass $m_a$. It turns out that the sensitivity has a sharp peak at the resonance $m_{a} \simeq \omega_a$ for both coaxial and parallel designs. For the  optimistic signal bandwidth $\Delta \nu \simeq 1/t$ regarding expected limit is estimated at the level of
$ g_{a\gamma \gamma}\lesssim 5\times 10^{-11}\, \mbox{GeV}^{-1}$ for $m_{a}  \lesssim \omega_{a}/2$. 

The detection  of a relatively small signal in a cavity with high intensity pump mode may lead to challenging
technical issues, mainly related to filtering of the tiny signal mode from the very intensive pump mode. 
To resolve this issue, one may consider a  ``bottle-shape''  cavity geometry  similar to that 
discussed in~\cite{Bogorad:2019pbu}. 

\vspace{0.3 cm}

\noindent\textbf{MM emitter + M$^*$F receiver.}
\begin{figure*}[t!]\centering
\includegraphics[width=0.4\linewidth]{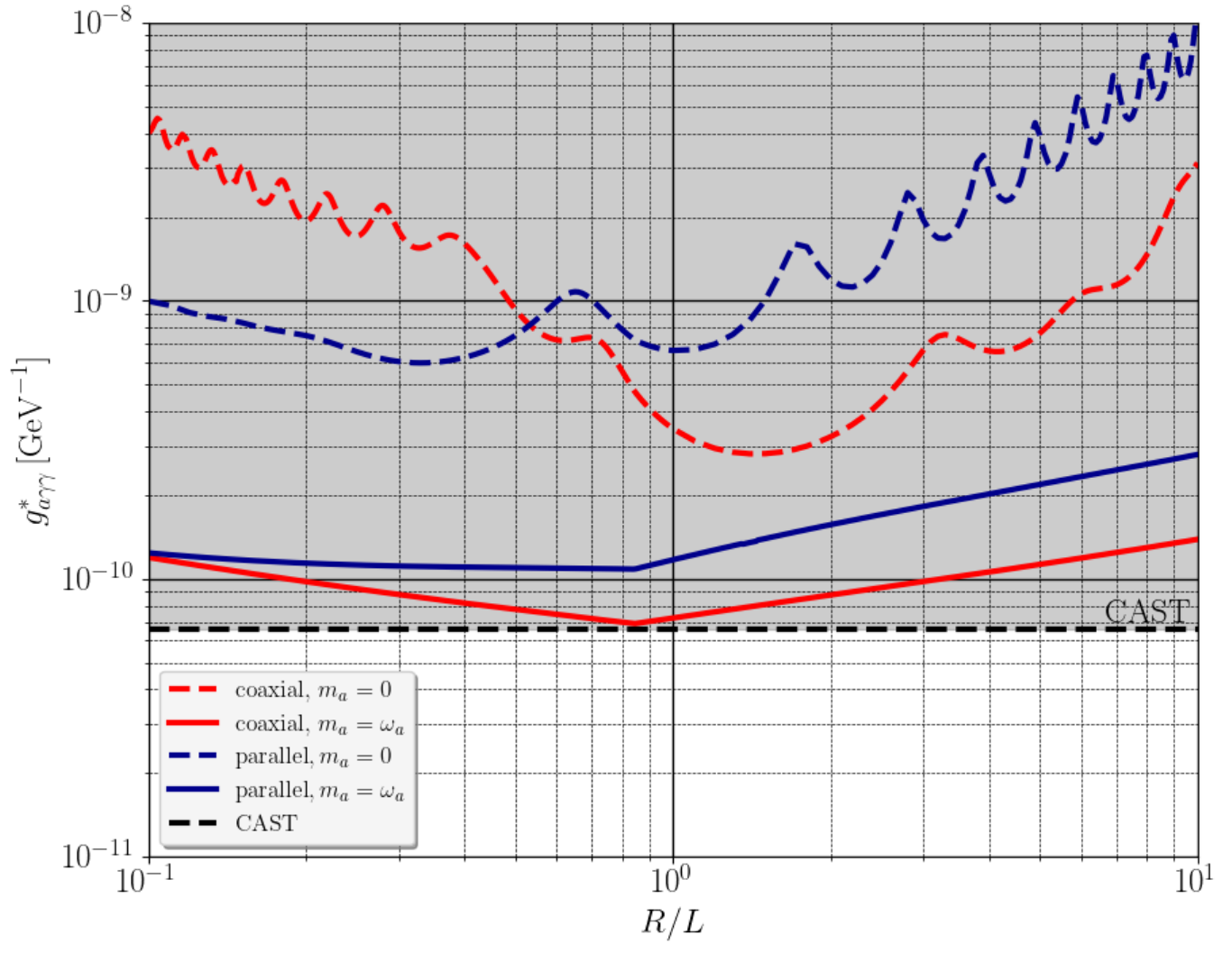}
\includegraphics[width=0.4\linewidth]{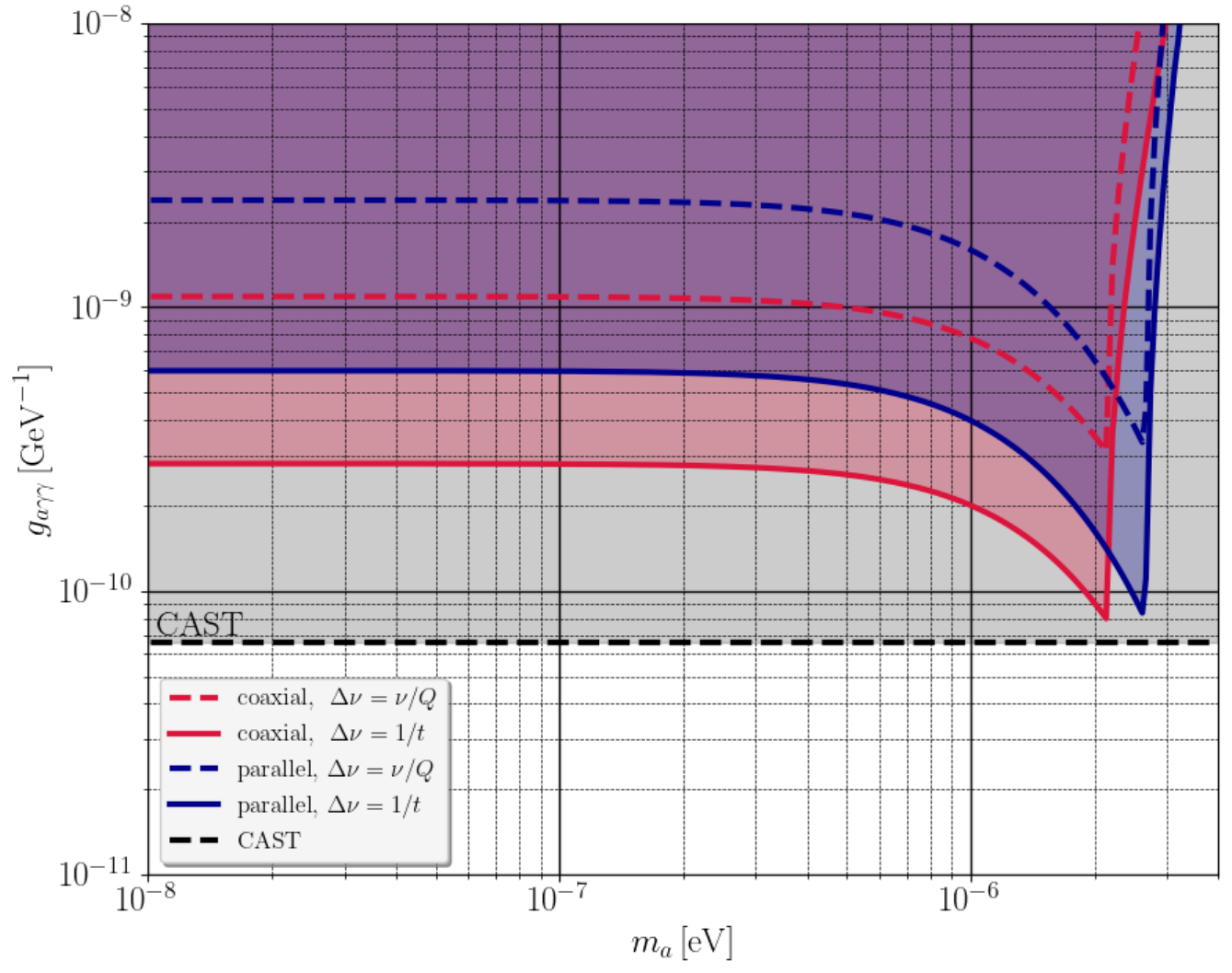}
\caption{Fig.~5: The sensitivity of the {\bf MM emitter + M$^*$F receiver} setup for  TM$_{010}$ + TE$_{011}$ 
emitter pump modes  and TM$_{010}$ detection signal mode.  The case of ALPs frequency $\omega_a = \omega_+$ is considered.
Left panel: the sensitivity dependence on emitter cavity 
radius-to-length ratio $R/L$ (fixed volume of $V_{\rm em} = 1 \, {\rm m}^3$ and 
fixed length of $L_{\rm rec} = 0.5 \, \mbox{\rm m}$). 
Right panel: sensitivity as a function of ALPs mass  at optimal $R/L$  for coaxial ($R/L \simeq 1.44$, $R \simeq 0.77 \, {\rm m}$, 
$L \simeq 0.54 \, {\rm m}$, $R_{\rm rec} \simeq 0.22 \, {\rm m}$, $L_{\rm rec} \simeq 0.5 \, {\rm m}$) and 
parallel ($R/L \simeq 0.33$, $R \simeq 0.47 \, {\rm m}$, $L \simeq 1.43 \, {\rm m}$, $R_{\rm rec} \simeq 0.18 \, {\rm m}$, 
$L_{\rm rec} \simeq 0.5 \, {\rm m}$) design.  The distance between cavity walls  is $\Delta = 0.5$ m, the volume of the emitter 
cavity is $V_{\rm em}=1$ m$^3$.  Integration time is $t = 10^6 \, {\rm s}.$ The temperature of the receiver is taken as $T=1.5$ K.}
\label{fig:Salnikov}
\end{figure*}
The next  setup that we consider in our study  consists of a production SRF cavity with two pump modes and a detection RF cavity 
immersed  into static magnetic field \cite{Salnikov:2020urr, Salnikov:2022rmi}. 

In Fig.~\ref{fig:Salnikov} we show the sensitivity of this type of experiment for the characteristic volume of the emitter cavity $V_{\rm em} \simeq 1\, \mbox{m}^3$ and its quality factor  $Q_{\rm em} \simeq 10^{10}$. Amplitudes of the emitter pump modes are   $B_0^{\rm em} = 0.1 \, {\rm T} \ (E_0^{\rm em} = 30 \, {\rm MV/m})$ to avoid  destruction of the superconducting state. The expected 
 power of the emitter cavity is $P_{\rm em} \simeq 0.1 \, {\rm kW}$. The distance between receiver and emitter walls is $\Delta = 0.5$ m. The pump modes of the emitter are TM$_{010}$ and TE$_{011}$, and the signal mode of the receiver is TM$_{010}$. Note that the receiver  must be smaller than the emitter
for the frequency equality $\omega_1 + \omega_2 = \omega_s$ to be fulfilled. The receiver quality factor is 
 $Q_{\rm rec} \simeq 10^{5}$  and the typical value of the static magnetic field 
 $B_0^{\rm rec} = 3 \, {\rm T}$.  

In Fig.~\ref{fig:Salnikov} (left panel) we show the typical expected reach for this setup as a function of $R/L$ for the emitter cavity.  We emphasize  that the regarding bounds are ruled out by the CAST facility at $g_{a\gamma\gamma} \lesssim 3.0 \times 10^{-10}\, \mbox{GeV}^{-1}$. This can be also justified from the right panel of Fig.~\ref{fig:Salnikov}  where the typical bounds are shown in the $(g_{a\gamma\gamma}, m_{a})$ plane. 


\vspace{0.3 cm}

\noindent\textbf{MF emitter + M$^*$M receiver.}
\begin{figure*}[t!]\centering
\includegraphics[width=0.4\linewidth]{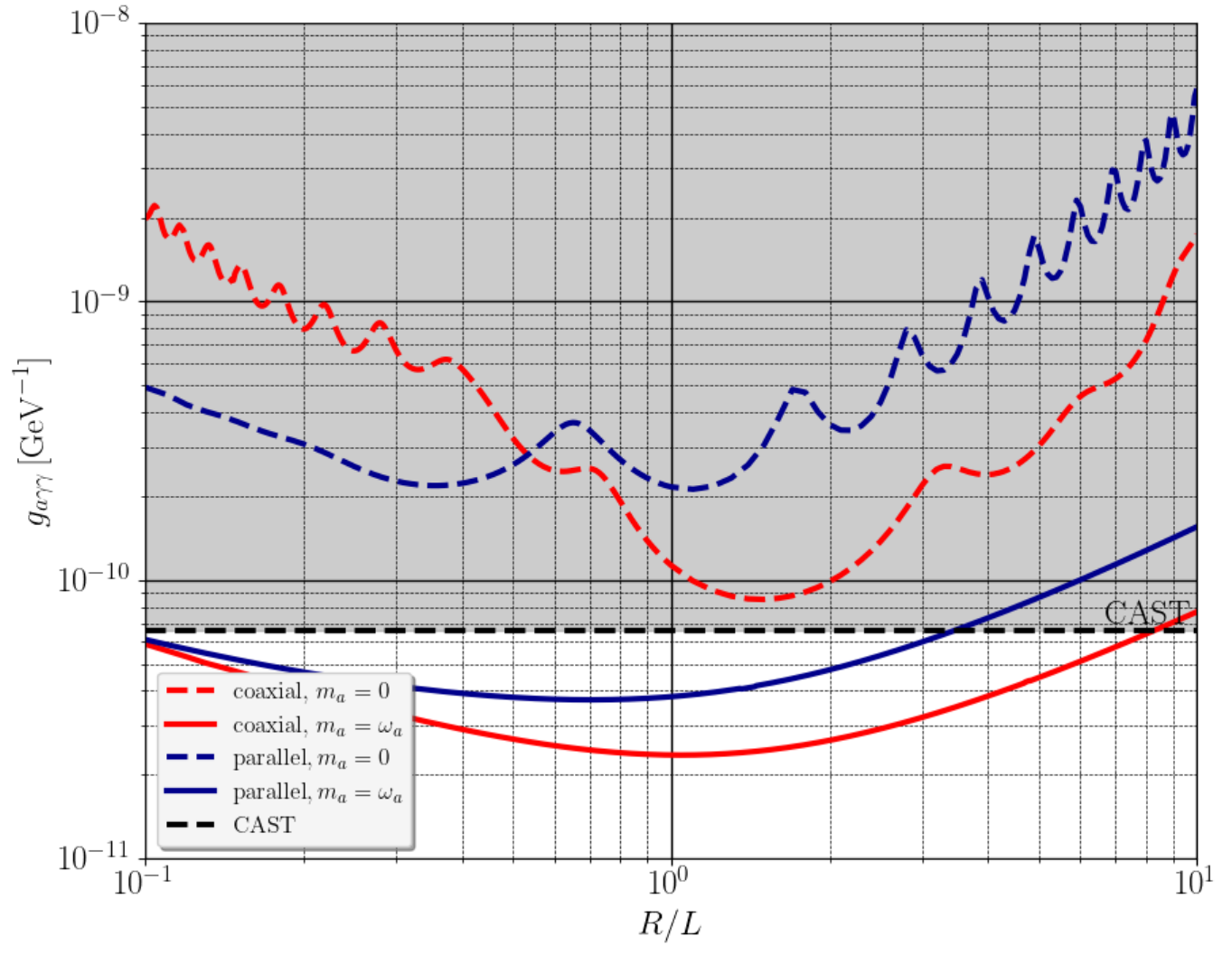}
\includegraphics[width=0.4\linewidth]{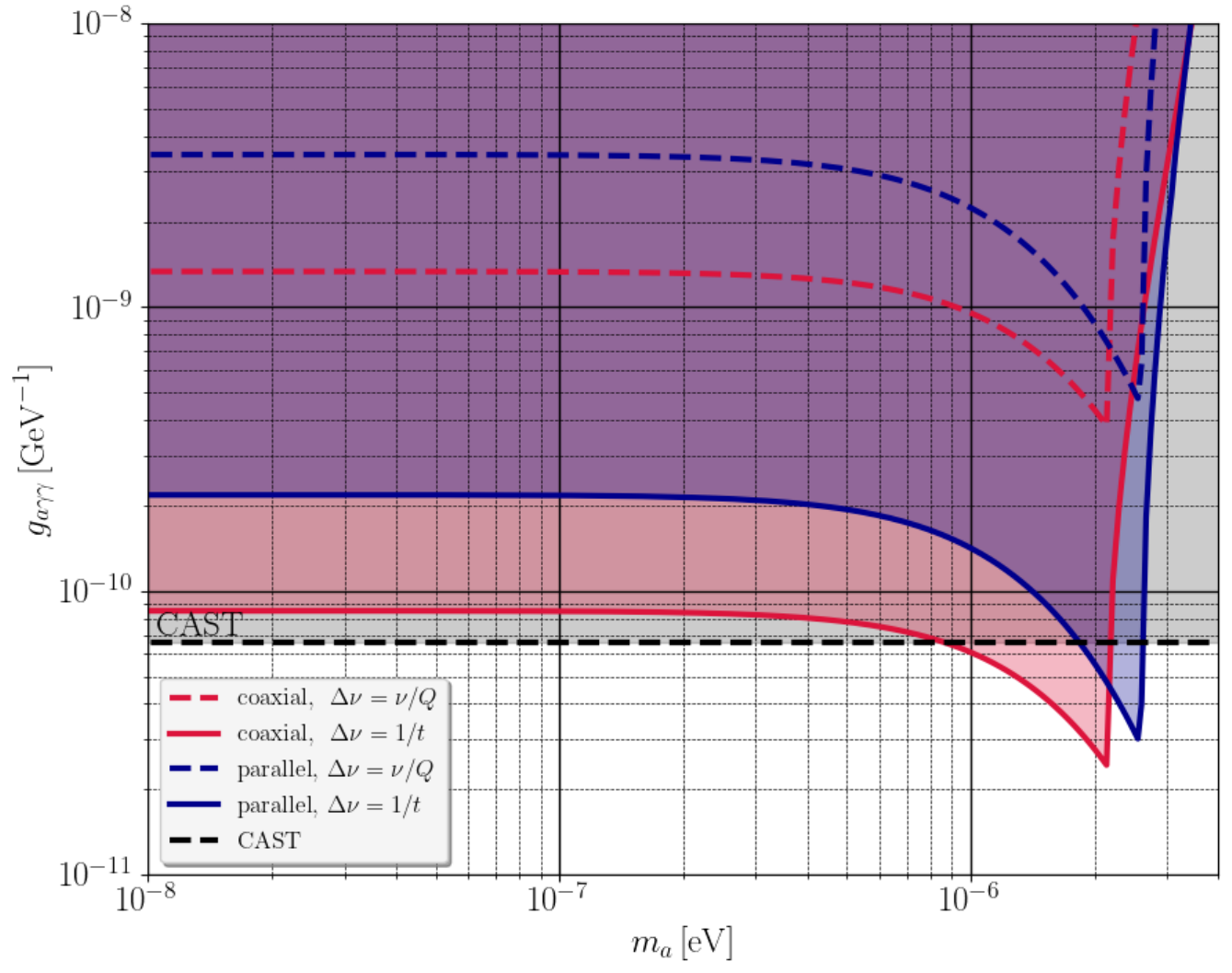}
\caption{Fig.~6: The sensitivity of the {\bf MF emitter + M$^*$M receiver} setup for  TM$_{010}$ emitter pump mode  and TM$_{010}$ receiver pump mode and TE$_{011}$  receiver signal mode. 
Left panel: the sensitivity dependence on receiver cavity 
radius-to-length ratio $R/L$ (fixed volume of $V_{\rm rec} = 1 \, {\rm m}^3$ and 
fixed length of $L_{\rm em} = 0.5 \, \mbox{\rm m}$). 
Right panel: sensitivity as a function of ALPs mass  at optimal $R/L$  for coaxial ($R/L \simeq 1.46$, $R \simeq 0.78 \, {\rm m}$, 
$L \simeq 0.53 \, {\rm m}$, $R_{\rm em} \simeq 0.22 \, {\rm m}$, $L_{\rm em} \simeq 0.5 \, {\rm m}$) and 
parallel ($R/L \simeq 0.36$, $R \simeq 0.49 \, {\rm m}$, $L \simeq 1.35 \, {\rm m}$, $R_{\rm em} \simeq 0.18 \, {\rm m}$, 
$L_{\rm em} \simeq 0.5 \, {\rm m}$) design.  The distance between cavity walls  is $\Delta = 0.5$ m, the volume of the receiver
cavity is $V_{\rm rec}=1$ m$^3$.  Integration time is $t = 10^6 \, {\rm s}.$ The temperature of the receiver is taken as $T=1.5$ K.}.
\label{fig:6}
\end{figure*}
The final  setup  consists of a production RF cavity with a pump mode  into static magnetic field  and a detection SRF cavity with a pump mode.

\begin{table*}[h!]
    \centering
    \begin{tabular}{|l|c|c|c|c|c|c|c|}
    \hline
    \hline
    \multicolumn{1}{|c|}{Type of the setup} & $B^{\rm em, (1)}_0$ & $B^{\rm em, (2)}_0$ & $B^{\rm rec}_0$ & $Q_{\rm rec}$ & $P_{\rm em}$ & $|\cal G|$ & $g_{a\gamma\gamma}$ \\
    \hline
     MF \hspace{1mm}em. + M$^*$F \hspace{1mm}rec. &$0.01 \, \rm T$ & $3 \, \rm T$ & $3 \, \rm T$ & $10^5$ & $100 \, \rm kW$ & $10^{-2}$ & $3\times 10^{-11} \,  {\rm GeV}^{-1}$ \\
    \hline
     MM em. + M$^*$M rec.&  $0.1 \, \rm T$ & $0.1 \, \rm T$ & $0.1 \, \rm T$ & $10^{10}$ & $0.1 \, \rm kW$ & $10^{-3}$ & $5\times 10^{-11} \,  {\rm GeV}^{-1}$ \\
    \hline
     MM em. + M$^*$F \hspace{1.2mm}rec. &  $0.1 \, \rm T$ & $0.1 \, \rm T$ & $3 \, \rm T$ & $10^{5}$ & $0.1 \, \rm kW$ & $10^{-3}$ & $3\times 10^{-10} \, {\rm GeV}^{-1}$ \\
    \hline
    MF \hspace{1mm}em. +  M$^*$M rec. &  $0.01 \, \rm T$ & $3 \, \rm T$ & $0.1 \, \rm T$ & $10^{10}$ & $100 \, \rm kW$ & $10^{-3}$ & $9\times 10^{-11} \, {\rm GeV}^{-1}$ \\
    \hline
    \hline
    \end{tabular}
    \caption{Table~1: Comparison of the characteristics for various experimental setups. The geometrical formfactor $|{\cal G}|$ and the setup sensitivity $g_{a\gamma\gamma}$ are presented for the best ratio of $R/L$ of coaxial location and the mass of ALPs $m_a \lesssim \omega_a/2$.}
    \label{tab:my_label}
\end{table*}

In Fig.~\ref{fig:6} we show the sensitivity of this type of experiment for the characteristic volume of the receiver cavity $V_{\rm rec} \simeq 1\, \mbox{m}^3$ and its quality factor  $Q_{\rm rec} \simeq 10^{10}$. The amplitude of the emitter pump mode is  $E_0^{\rm em} = 3 \, {\rm MV/m} \ (B_0^{\rm em} = 0.01 \, {\rm T})$, the magnitude of static magnetic field if $B_{\rm ext} = 3\, {\rm T}$. The expected 
 power of the emitter cavity is $P_{\rm em} \sim 100 \, {\rm kW}$. The distance between receiver and emitter walls is $\Delta = 0.5$ m. The pump mode of the emitter is TM$_{010}$, the pump mode of the receiver is TM$_{010}$ and the signal mode of the receiver is TE$_{011}$. Note that the emitter  must be smaller than the receiver
for the frequency equality $\omega_1 + \omega_2 = \omega_{\rm em}$ to be fulfilled. The receiver quality factor is 
 $Q_{\rm rec} \simeq 10^{10}$  and the typical value of the pump mode amplitude is
 $B_0^{\rm rec} = 0.1 \, {\rm T}$.  

In Fig.~\ref{fig:6} (left panel) we show the typical expected reach for this setup as a function of $R/L$ for the receiver cavity.  We emphasize  that the regarding bounds are ruled out by the CAST facility at $g_{a\gamma\gamma} \lesssim 9.0 \times 10^{-11}\, \mbox{GeV}^{-1}$ for the mass range $m_a\lesssim \omega_a/2$. This can be also justified from the right panel of Fig.~\ref{fig:6}  where the typical bounds are shown in the $(g_{a\gamma\gamma}, m_{a})$ plane. Remarkably however
that the typical peak bounds at $m_a \simeq 2\times 10^{-6}\, \mbox{eV}$ can rule out the CAST limits.

\vspace{0.3 cm}

\noindent\textbf{Results and discussion.}
We compared four types of the LSW radio setups for ALP searches and determined the best design for them. We  summarize our study presenting important parameters for each setup in Table~1.

We concluded that the MF emitter + M$^*$F receiver and the MM  emitter + M$^*$M  receiver setups
can achieve the similar top sensitivity $g_{a\gamma\gamma}\lesssim (3 - 5) \times 10^{-11}\, \mbox{GeV}^{-1}$  
at  $m_a \lesssim \omega_a/2$. In particular, it turns out that the larger electromagnetic field combination and the 
geometrical formfactor of RF cavities  compensate its smaller quality factor.     

Moreover, we find that the best relative location of the cavities is 
coaxial with the ratio of $R/L \simeq 1.6$.

The MF emitter + M$^*$F receiver setup is a modification of the CROWS experiment \cite{Betz:2013dza} that implies larger volume of the cavities $V_{\rm em} \simeq V_{\rm rec}\simeq 1\, \mbox{m}^3$, lower temperature, and narrower bandwidth of the signal, $\Delta \nu \simeq 1/t$.  
However, there is a disadvantage of this setup that implies the relatively large emitter 
power $P_{\rm em} \sim 100 \, {\rm kW}$.

The advantage of the MM emitter +  M$^*$M receiver setup is that its emitter power is 4 orders of magnitude 
smaller than  the previous one.

However, in this case the main technical challenges would be related to the signal mode filtering from the pump mode and fine tuning of cavity sizes. 

Given the benchmark parameters, the last two setups, MM emitter + M$^*$F receiver and MF emitter + M$^*$M receiver, has the weakest sensitivity, see Table~\ref{tab:my_label}.
Moreover, the typical bounds $g_{a\gamma\gamma}\lesssim \mathcal{O}(10^{-10})\, \mbox{GeV}^{-1}$  would be ruled out by the CAST. Also, there is a disadvantage of these proposals.
In particular, the condition 
$\omega_1 + \omega_2 = \omega_s$ implies the specific type of the emitter modes,  the latter is  linked to the sizes of the cavity.  Moreover, the modification  of the pump modes would require the changing of the 
receiver  geometry. The disadvantages of the MF emitter + M$^*$M receiver  include also technical difficulties of the first two setups.

\paragraph{Acknowledgements.}
The authors thank Sergey Troitsky, Yury Senichev, and Yonatan Kahn for  
valuable discussions and helpful suggestions. The work on optimal geometry of the 
 setups  is supported by  RSF grant 21-72-10151. The work of DS on design comparison of the setups is 
 supported by the BASIS Foundation, grant no. 22-2-1-17-1.

\bibliographystyle{h-physrev}
\bibliography{grb}
\end{document}